\documentclass[%
 reprint,
superscriptaddress,
 amsmath,amssymb,
pra,
]{revtex4-2}

\usepackage{graphicx}
\usepackage{dcolumn}
\usepackage{bm}

\usepackage[bookmarks=true,colorlinks=true,urlcolor=blue,linkcolor=blue,citecolor=blue,breaklinks=true]{hyperref}
\usepackage{graphicx}
\usepackage[table]{xcolor}       

\usepackage{lipsum}

\newcommand{\UBX}[0]{Université de Bordeaux, 351 Cours de la Libération, 33405 Talence, France}

\newcommand{\UPC}[0]{Departament de F\'{i}sica, Universitat Polit\`{e}cnica de Catalunya, Campus Nord B4-B5, E-08034, Barcelona, Spain}

\begin{document}


\title{A quantum Monte Carlo based functional for Dysprosium dipolar systems}

\author{Raúl Bombín}
\email{raul.bombin@ehu.eus}
\affiliation{\UBX}
\affiliation{\UPC}

\author{Viktor Cikojevi{\'c}}
\email{cikojevic.viktor@gmail.com}
 \affiliation{Split, Croatia}

\author{Ferran Mazzanti}
\affiliation{\UPC}
\author{Jordi Boronat}
\affiliation{\UPC}

\date{\today}

\begin{abstract}

We present a quantum Monte Carlo based density functional to describe droplet formation and supersolidity in dipolar systems.
The usual Lee-Huang-Yang term, accounting for quantum correlations in the conventional extended Gross-Pitaevskii equation (eGPE), has been substituted by the correlation energy evaluated with Quantum Monte Carlo. We demonstrate the ability of the new functional to reproduce existing experimental data for the minimum critical number of atoms $N_\mathrm{c}$ required for droplet formation. $N_\mathrm{c}$ is a challenging quantity for theoretical predictions, and the eGPE provides only a qualitative description of it, mainly when it is applied to Dysprosium. We also use the new approach to characterize the BEC-supersolid transition. The quantum Monte Carlo based functional can be easily implemented in any existing eGPE code, improving the description of dipolar systems without increasing the computational cost.

\end{abstract}

\maketitle


\section{Introduction}

Ultracold dipolar systems made up of lanthanide atoms, mainly Dy~\cite{Lu2011}, Er~\cite{Aikawa2012}, or their mixtures~\cite{Trautmant2018}, have been extensively studied in recent years (see Ref~\cite{Chomaz2023} for a experimental review). The long-range and anisotropic nature of the dipolar interaction gives place to unique many-body phenomena such as the formation of ultradilute quantum droplets~\cite{Schmitt2016,Chomaz2016,Ferrierbarbut2016}, the existence of supersolid phases~\cite{Chomaz2019,Tanzi2019a,Bottcher2019}, and recently, the observation of anomalous temperature behavior~\cite{sohmen2021,Sanchez-Baena2023}.
State-of-the-art experiments provide unprecedented insight into this correlated regime. Some achievements in this respect are the measurement of the roton of the quasi-particle excitation spectrum~\cite{Chomaz2018}, of the collective excitations~\cite{Barbut2018,Tanzi2021,Tanzi2019b}, and of the static structure factor, which has been used to characterize density fluctuations across the BEC-supersolid transition~\cite{Schmidt2021,Hertcorn2021}. Recently it has been claimed the observation of the first BEC of dipolar molecules~~\cite{bigagli2023cond} what would potentially give access to the study new phenomena on dipolar systems.

The extended Gross-Pitaevskii equation (eGPE), which includes beyond mean-field contributions in the Lee-Huang-Yang (LHY) form~\cite{schutzhold2006,Lima2011,Lima2012}, has been extensively utilized to describe ultracold dipolar systems with remarkable success~\cite{Baillie2016,Bisset2016,Wachtler2016,Saito2016,Zhang2019,Blakie2020,Bisset2021}. For instance, in Er systems, this theory allows to describe the formation of droplets~\cite{Chomaz2016}, the supersolid phases~\cite{Natale2019}, and also the excitation spectrum~\cite{Chomaz2018}. Similarly, for Dy atoms, the eGPE has been successfully applied to evaluate the superfluid critical velocities~\cite{Wenzel2018}, collective excitations~\cite{Barbut2018,Tanzi2021}, and to describe striped states in oblate traps~\cite{Wenzel2017}.
Despite these successes, there are still some challenges that need to be addressed. It has been shown that incorporating the LHY correction does not always result in an improvement, for instance in the description of the roton minimum for Er systems~\cite{Petter2019}. More significantly, eGPE prediction of the critical (minimum) atom number to form a droplet in Dy systems is only qualitative~\cite{Bottcher2019crit}. Besides these evidences, it is worth highlighting a fundamental drawback that affects the eGPE description in the regime where dipolar interaction dominates, namely its dynamic instability in the long wavelength limit.

Efforts to refine the eGPE description of ultracold dipolar systems have attempted to improve the description of quantum correlations~\cite{Cherny2019,Boudjemaa2020,Zhang2022} or to include thermal effects in the model~\cite{Boudjemaa2016,Boudjema2017}.
An alternative approach consists of using quantum  Monte Carlo (QMC), which evaluates exactly quantum correlations at all orders.
However, its huge computational cost hinders the direct calculation of large systems, containing a number of atoms $N>10^4$.
To tackle this limitation, models in which the experimental scattering length is rescaled to allow for the formation of small droplets have been used~\cite{Macia2016,Bombin2023}.
QMC calculations have also estimated the critical atom number for a small droplet ($N\sim10^3$)~\cite{Bottcher2019crit}. The critical atom number arises from a subtle balance between dipolar attraction and quantum correlations, making it a challenging quantity to be calculated.  In the context of Bose-Bose mixtures, it has been shown that it is possible to avoid the QMC limitation in the number of atoms by constructing a density functional~\cite{cikojevik2020,cikojevic2020_2,cikojevic2021,Kopicinski2023} in the same approach that has been used previously in the study of liquid $^4$He drops~\cite{Ancilotto2017,Barranco2006}. Such a functional improves the description of quantum correlations, as it includes information from the QMC equation of state (EOS). In this work, we present a QMC based density functional for dipolar systems. We specialize it to the study of Dy (Dy-DF). The Dy-DF is able to accurately reproduce the critical atom number for droplet formation, highlighting the potential of this approach for advancing our understanding of ultracold dipolar droplets. We also show that it can be applied to study other relevant phenomena in Dy systems, such as the BEC-supersolid transition.

The paper is organized as follows. In section~\ref{subsec:system} we provide a description of the system that we study. In sections~\ref{subsec:egpe} and \ref{subsec:pigs} the eGPE and path integral ground state (PIGS) methods are briefly described. 
PIGS results for the equation of state (EOS) are discussed in Secs.~\ref{sec:EOS}.
In Sec.~\ref{sec:densityfunc} we explain how do we use the  EOS obtained with PIGS method to construct an eGPE like density functional, that we call Dysprosium  density functional (Dy-DF). The accuracy of the Dy-DF is benchmarked using the available experimental data for the critical atom number of $^{162}$Dy and $^{164}$Dy droplets~\cite{Bottcher2019crit} in section~\ref{sec:benchmarck} .
In Sec.~\ref{sec:BEC-supers-trans} we use the Dy-DF to study the BEC-supersolid transition paying attention to the different observables. We also discuss the differences that emerge between the eGPE predictions and our theory. Finally in Sec.~\ref{sec:conclusions} some conclusions are summarized.

\section{Method}
\label{sec:method}
\subsection{The dipolar system}
\label{subsec:system}
We study  a system of $N$ $^{162}$Dy magnetic atoms with  all the magnetic moments aligned along the $Z$ direction in space.  
Such a system can be described by the following Hamiltonian
\begin{equation}
\hat{H}= \hat{T} +  \hat{V}_{\mathrm{trap}} +  \hat{V}_{\mathrm{2B}} ,
    \label{eq:Hamiltonian}
\end{equation}
where $\hat{T}=-\sum_{i=1}^N\frac{\hbar^2\hat{\nabla}_i^2}{2M}$ is the quantum
kinetic energy operator with $M$  the mass of $^{162}$Dy,  $\hat{V}_{\rm trap}$ 
is an external trapping potential, $\hat{V}_\mathrm{2B} = 
\hat{V}_{\mathrm{SR}}+\hat{V}_{\mathrm{dd}}$ the two-body potential, consisting 
on a short-ranged Lennard-Jones $\hat{V}_{\mathrm{SR}}$,  and a dipolar 
one $\hat{V}_{\mathrm{dd}}$,
\begin{align}
V_{\mathrm{SR}}(\mathbf{R}) &= \sum_{i<j}^N\frac{C_{12}}{r_{ij}^{12}}-\frac{C_6}{r_{ij}^6},\\
V_{\mathrm{dd}}(\mathbf{R}) &= \sum_{i<j}^N\frac{C_{\mathrm{dd}}}{4\pi}\frac{1-3\cos^2\theta_{ij}}{r_{ij}^3},
\label{eq:2Bpots}
\end{align}
with $\mathbf{R}=(\mathbf{r}_1,\mathbf{r}_2,...,\mathbf{r}_N)$ the set of 3N-coordinates, 
$\mathbf{r}_i$ the position of the ith particle, $\mathbf{r}_{ij}=\mathbf{r}_i-\mathbf{r}_j$,  
$(r_{ij},\theta_{ij})$ the polar coordinates of the vector $\mathbf{r}_{ij}$ 
and the ${C_i}$ constants determine the strength of the different contributions to $V_{\mathrm{2B}}$. 
The $C_{12}$ value is chosen so that the total interaction potential has the desired s-wave scattering length $a_s$, by solving the Lippmann-Schwinger equation associated with the T matrix of the full interaction~\cite{Bottcher2019crit}, 
the van der Waals $C_6$ coefficient is fixed to the Dysprosium experimental value~\cite{Li2017},
and $C_{\mathrm{dd}}=\mu_0m^2$ with $\mu_0$ the vacuum magnetic permeability and 
$m\approx 10~\mu_B$ the magnetic moment of $^{162}$Dy.
The trapping potential $\hat{V}_{\mathrm{trap}}$  is chosen as a one-body 
harmonic potential. To evaluate properties of a bulk, infinite system, the 
external potential is set to zero. 
The results presented in this work are in units of the dipolar length $r_0 = 
\frac{C_{dd} M}{4\pi\hbar^2}$ and the corresponding dipolar unit of energy 
$\mathcal{E}_0= \frac{\hbar^2}{Mr_0^2}$, unless otherwise is stated.

\subsection{The Gross Pitaevskii equation}
\label{subsec:egpe}

The Gross-Pitaevskii equation  is a non-linear differential equation that allows to describe a bosonic system at zero temperature. It provides a mean-field description assuming that almost all the system remains in the condensate. 
When quantum correlations in the LHY form are included it is usually referred as extended Gross-Pitaevskii equation (eGPE), that in the case of a dipolar system reads~\cite{Bisset2016,Wachtler2016,Saito2016,Lima2011,Lima2012}
\begin{equation}
	i \hbar \frac{\partial}{\partial t} \Psi(\textbf{r}, t)= \mu \Psi(\textbf{r}, t),
\label{eqn:eGPE}
\end{equation}
where $\Psi$ is the one-body mean-field wave-function and $\mu$ the chemical potential that reads
\begin{equation}
    \mu = -\frac{\hbar^{2} \nabla^{2}}{2 M}+V_{\mathrm{\mathrm{trap}}}+g|\Psi|^{2} +\Gamma_{\mathrm{QF}}|\Psi|^3 + \Phi_{dd},
\label{eqn:mu}  
\end{equation}
with $g = 4\pi \hbar^2 a_s / M$  the  coupling constant, the fourth term 
$\Gamma_{\mathrm{QF}}|\Psi|^3$ is the LHY correction including quantum 
correlations~\cite{schutzhold2006,Lima2011,Lima2012},
\begin{equation}
\Gamma_{\mathrm{QF}}|\Psi|^3=\frac{32 g \sqrt{a_s^3}}{3 \sqrt{\pi}} \mathcal{Q}_{5}\left(\varepsilon_{\mathrm{dd}}\right)|\Psi|^3,
\label{eqn:LHY}
\end{equation}
with $\varepsilon_{\mathrm{dd}} = a_{\mathrm{dd}} / a_s$, $a_{\mathrm{dd}} = r_0 / 3 =  \frac{C_dd M}{12\pi\hbar^2}$, $Q_5(\varepsilon_{\mathrm{dd}}) = \dfrac{1}{2} \displaystyle \int_{0}^{\pi} \mathrm{d} \alpha \sin\alpha \left[1 + \varepsilon_{\rm dd} (3 \cos^2 \alpha - 1)\right]^{5/2}$ and the last term in Eq.~\eqref{eqn:mu} is a non-local term that accounts for the dipolar interaction
\begin{equation}
   \Phi_{dd}(\textbf{r}) = \int d\textbf{r}^\prime V_{\mathrm{dd}}(\textbf{r}-\textbf{r}^\prime)|\Psi(\textbf{r}^\prime,t)|^2.
\end{equation}

At zero temperature, the chemical potential of Eq.~\eqref{eqn:eGPE} is related to the internal energy $U$ through the thermodynamic relation  
\begin{equation}
\mu = \left(\frac{\partial U}{\partial N}\right)_V = \frac{\rho}{N}\left(\frac{\partial U}{\partial \rho}\right)_V = \left(\frac{\partial \mathcal{E}}{\partial \rho}\right)_V,
\label{eqn:chempot}
\end{equation}
 where $V=N/\rho$ is the volume of the system, $\rho$ is the density and $\mathcal{E}=\rho U/N$ is the energy per unit of volume or the energy density. 

\subsection{Path Integral Ground state}
\label{subsec:pigs}

PIGS is an stochastic method that allows to evaluate properties of quantum many-body systems at zero temperature~\cite{Ceperley1995,Sarsa2000,Galli2003,Cuervo2005}. 
In the case of bosonic systems it provides exact results within  some statistical uncertainty. 
The method allows to sample the ground state of the system $\phi_0$ by propagating in imaginary time, $\tau$, a many-body trial wave function $\Phi_\mathrm{T}(\textbf{R},\tau)$
\begin{equation}
\phi_0 = \lim_{\tau\to\infty}\Phi(\textbf{R},\tau) = \int\lim_{\tau\to\infty}d\textbf{R}^\prime G(\textbf{R},\textbf{R}^\prime;\tau)\Phi_T(\textbf{R}^\prime,0),
    \label{eqn:pigs-prop}
\end{equation}
where $G(\textbf{R},\textbf{R}^\prime;\tau)$ is the propagator operator 
in 
the imaginary-time interval $\tau$.

In general, the many-body propagator is unknown and instead one uses a 
short-time approximation $\delta\tau$  and iterates Eq. \ref{eqn:pigs-prop}. 
To achieve large imaginary times, a set of $M_b$ 
intermediate coordinates (beads), $\{\textbf{R}_i\}$, are introduced in 
Eq~\eqref{eqn:pigs-prop}, that takes the form
\begin{align}
 \phi_0 (\mathrm{R}_{N_b+1})= & \lim \limits_{\substack{%
    \delta\tau \to 0\\
    N_b \to \infty\\
    \tau \to \infty}} \int\prod_{i=1}^{M_b}d\textbf{R}_1d\textbf{R}_2...d\textbf{R}_{N_b}\nonumber\\
 &\times G(\textbf{R}_{i+1},\textbf{R}_i;\delta\tau)\Phi_\mathrm{T}(\textbf{R}_1,0) .
 \label{eqn:pigs-prop2}
\end{align}
The method guarantees that the ground state is reached for sufficiently small $\delta\tau$ and sufficiently large $N_b$ with $\tau = \delta\tau M_b\to\infty$, as long as the trial wave function is not orthogonal to the actual ground state of the system. We use a fourth order Chin action propagator~\cite{Chin2004,Chin2006} and permutations are sampled with the efficient worm algorithm~\cite{prokofev1998,Boninsegni2006}.  In the calculations presented in this work the trial wave function is chosen as a constant, which has been demonstrated to be enough even for correlated systems such as $^4$He~\cite{Rota2010}.

\begin{table}[t]
    \caption{PIGS results for the equilibrium density $\rho_\mathrm{eq}$ and binding energy per atom $E_\mathrm{b}$ at that density for a bulk $^{162}$Dy system.}
    \label{table:eq-dens}
\begin{ruledtabular}
\begin{tabular}{ c c c  }
$a_s/a_B$& $\rho_\mathrm{eq}r_0^2$ & $E_\mathrm{b}/\mathcal{E}_0$\\
\hline
60.00 & 0.068& -0.045\\
70.00 & 0.051& -0.033\\
80.00 & 0.039& -0.024\\
90.00 & 0.029& -0.018\\
101.57 & 0.023& -0.013\\
110.00 & 0.018& -0.011\\
\end{tabular}
\end{ruledtabular}
\end{table}

\begin{figure*}[t!]
    \centering
    \includegraphics[width=1.0\linewidth]{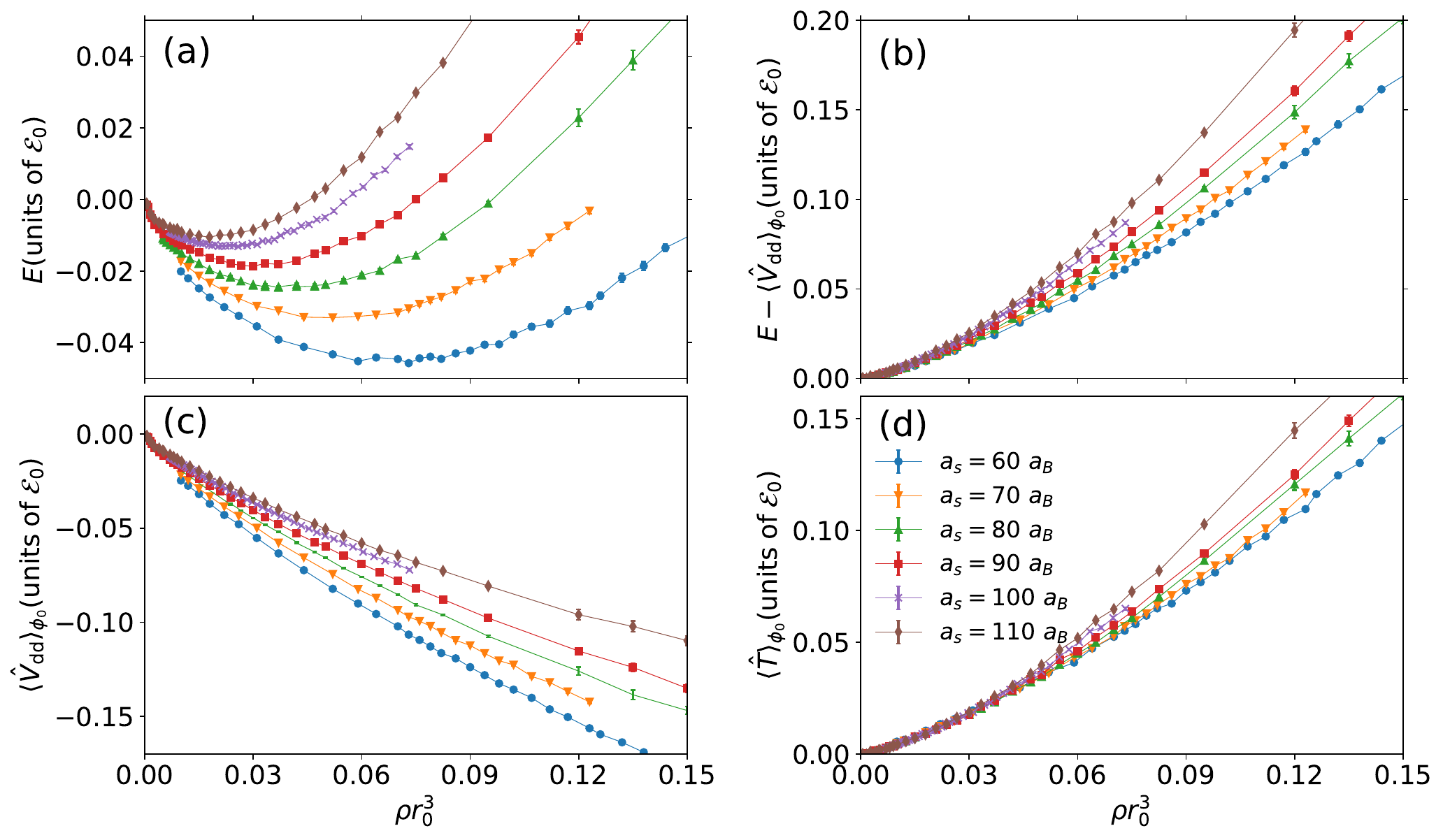}
    \caption{PIGS results for a)energy per particle, $E$,  b)energy per particle without dipolar contribution $E -  \left<\hat{V}_\mathrm{dd} \right>_{\phi_0}$, c)dipolar energy per particle $\left<\hat{V}_\mathrm{dd} \right>_{\phi_0}$, and d)correlation energy per particle $\left<\hat{T} \right>_{\phi_0}$, as a function of the density $\rho r_0^3$. Results are shown for different values of the scattering length. Lines are guides to the eye.}
    \label{fig:EOS}
\end{figure*}

\section{Equation of state}
\label{sec:EOS}
The bulk dipolar system is simulated by considering $N=512$ $^{162}$Dy atoms in a cubic box of lenght  $L=\sqrt{N/\rho}$ with fixed density $\rho$ and periodic boundary conditions.
We consider values of the scattering length that cover the  whole range of state-of-the-art experiments with  $^{162}$Dy and $^{164}$Dy atoms: $a_\mathrm{s}=\{60,70,80,90,101.57,110\}~a_B$, with $a_\mathrm{B}$ the Bohr radius.  The energy per particle $E(\rho)$ of Fig.~\ref{fig:EOS}(a) corresponds to the EOS of a liquid \textit{i.e.} it exhibit a minimum in the energy per particle at the equilibrium density $\rho_\mathrm{eq}$. The values of the equilibrium density $\rho_\mathrm{eq}$ and the binding energy per particle $E_{\mathrm{b}}$ are shown in table~\ref{table:eq-dens}.  For small values of the scattering length the dipolar interaction dominates and the system forms a denser liquid with a larger binding energy. 
The formation of the liquid arises from the balance between dipolar attraction 
and the 
repulsion coming from  quantum correlations.

To give a further insight into the energetic balance of the dipolar system, in Fig.~\ref{fig:EOS}(b) the dipolar energy contribution to the energy per particle $\left<\hat{V}_\mathrm{dd} \right>_{\phi_0}$ is subtracted from $E$ and in  Fig.~\ref{fig:EOS}(c) the $\left<\hat{V}_\mathrm{dd} \right>_{\phi_0}$ is plotted. In the above expressions $\left<\hat{O}_\mathrm{dd} \right>_{\phi_0}$ means the expectation value of the operator $\hat{O}$ in the ground ground state of the system. Results in Fig.~\ref{fig:EOS}(b) and (c) show that the binding of the dipolar systems is entirely caused by the dipolar attraction. In this respect, it is worth remarking that $E$ arises from large cancellation between these two terms, what hints to the subtle energetic balance that exists between short range repulsion, quantum correlation energy and dipolar interaction in the ultracold regime. 
To elucidate the role that quantum correlations play, in panel (d) of the same 
figure we show explicitly the kinetic (correlation) energy, $\left<\hat{T} 
\right>_{\phi_0}$. Comparing panels (b) and (d) it is clear that quantum 
correlations dominate the repulsive contribution to $E$ and thus, play a major 
role in the stabilization of the system. This is a well known fact in the 
ultracold gases community although an accurate quantitative evaluation of 
quantum correlations for realistic experimental parameters was lacking.
From the theoretical side, the inclusion of quantum correlations in the form of a LHY term in the Gross Pitaevskii equation allows to stabilize the droplets and has allowed to describe the mechanism of droplet formation (see for example Ref.~\cite{Baillie2016}). In the next section we construct a eGPE density functional by replacing the LHY term by the quantum correlations evaluated with PIGS.

\section{Dipolar density functional}
\label{sec:densityfunc}

The eGPE equation, that was introduced in sec.~\ref{subsec:egpe}, can be derived form the minimization of the following functional for the internal energy by imposing the usual energy variational principle $\delta(E-\mu N) = 0$~\cite{Ancilotto2018,cikojevic2020_2} 
\begin{align}
\label{eqn:MF-U}
U_{\rm LHY} = & \int \mathcal{E}_{\rm LHY} [\Psi] d\textbf{r} \\
 = & \int \left[\dfrac{\hbar^2}{2m}|\nabla \Psi|^2+ V_{\mathrm{\mathrm{trap}}}|\Psi|^2 +
 \frac{1}{2} g|\Psi|^4 +\dfrac{2}{5} \Gamma_{\rm QF} |\Psi|^5 \right.\nonumber\\
 &\left.+\frac{1}{2}\int d\textbf{r}^\prime V_{dd}(\textbf{r}-\textbf{r}^\prime)|\Psi(\textbf{r}^\prime,t)|^2|\Psi(\textbf{r},t)|^2 \right]d\textbf{r}\nonumber.
\end{align}
Using the local density approximation (LDA), that is, setting $\rho = |\Psi|^2$, one can rewrite the LHY-DF, $\mathcal{E}_{\rm LHY}$, as 
\begin{align}
\mathcal{E}_{\rm LHY} [\rho] 
 = &\dfrac{\hbar^2}{2m}|\nabla \sqrt{\rho}|^2+ V_{\mathrm{\mathrm{trap}}}\rho +
 \frac{1}{2} g\rho^2 +\dfrac{2}{5} \Gamma_{\rm QF} \rho^{5/2} \nonumber\\
 &+\frac{1}{2}\int d\textbf{r}^\prime V_{dd}(\textbf{r}-\textbf{r}^\prime)\rho(\textbf{r}^\prime)\rho(\textbf{r}) .
\label{eqn:MF-Efunc2}
\end{align}

Notice that in the case of a uniform bulk system [$\rho(r)\sim {\rm const}$, 
$V_{\mathrm{\mathrm{trap}}}=0$], the first two terms and the last one of 
Eq.~\eqref{eqn:MF-Efunc2} vanish and one obtains a functional for the energy 
density of the uniform dipolar bulk system that reads
\begin{align}
	\mathcal{E}_{\rm LHY}^{\rm BULK} [\rho]= &\frac{1}{2} g\rho^2 +\dfrac{2}{5} \Gamma_{\rm QF} \rho^{5/2} .
\label{eqn:MF-Efuncbulk}
\end{align}

On the other hand, using PIGS method, one can compute the EOS of the dipolar bulk system from the energy per particle $E_{\rm QMC}(\rho)=U_{\rm QMC}(\rho)/N$ for a given value of the s-wave scattering length. For the system described by the Hamiltonian of Eq.~\eqref{eq:Hamiltonian} it reads
\begin{equation}
    E_{\rm QMC}^{\rm BULK}(\rho = |\phi_0|^2) = 
    \left<\hat{T} \right>_{\phi_0} +
    \left<\hat{V}_\mathrm{SR} \right>_{\phi_0} +
    \left<\hat{V}_\mathrm{dd} \right>_{\phi_0}
    \label{eqn:QMC-EN}
\end{equation}
that is a sum of three contributions: correlation energy, short-range interaction and dipolar interaction.  
As it has been done in the context of helium~\cite{Barranco2006,Ancilotto2017} and Bose-Bose mixtures~\cite{cikojevik2020,cikojevic2021,cikojevic2020_2}, one could study the  non-uniform system, \textit{eg.} confined BEC, droplets..., under the LDA approximation with a functional of the form
\begin{align}
	\mathcal{E}_{\rm QMC} [\rho]= &\dfrac{\hbar^2}{2m}|\nabla \sqrt{\rho}|^2+ V_{\mathrm{\mathrm{trap}}}\rho +
 \mathcal{E}_{\rm QMC}^{\rm BULK} [\rho],
\label{eqn:QMC-Efunc}
\end{align}
with $\mathcal{E}_{\rm QMC}^{\rm BULK}=\rho E_{\rm QMC}^{\rm BULK}$. However, it is important to notice that the functional of Eq.~\eqref{eqn:QMC-Efunc} is isotropic, and thus it is not a good functional to describe dipolar systems, in which the anisotropy introduced by the dipolar interaction plays a major role. In the LHY functional the anisotropy is naturally included  in the non-local dipolar potential $V_{dd}$.

In this work we construct an eGPE like density functional for Dysprosium dipolar systems, the Dy-DF, replacing the quantum fluctuations term in Eq.~\eqref{eqn:MF-Efunc2} by the non-linear contribution to the quantum kinetic energy. To do so, we fit $\left<\hat{T}\right>_{\phi_0}( \rho)$ evaluated with PIGS with a function of the form of \eqref{eqn:MF-Efuncbulk}. Explicitly,
\begin{equation}
        \left<\hat{T}\right>_{\phi_0}= \alpha_{\rm QMC}\rho + \beta_{\rm QMC}\rho^{\gamma_{\rm QMC}}.
    \label{eqn:QMC-ekin-fit}
\end{equation}
This approach implies assuming that the linear term in the usual eGPE 
description 
is accurate enough and that only a modification in the quantum correlation 
term is needed in order to include beyond LHY effects. From a different 
approach, small modifications to the LHY-DF have been proposed in order to 
improve the accuracy in the description of quantum dipolar systems (see 
Ref.~\cite{Zhang2022}).
Due to the stochastic nature of the PIGS method, statistical uncertainties are present in our data.  Details of the fitting procedure  are given in appendix~\ref{app:PIGSfit} and the specific values of $\{\beta(a_\mathrm{s}),\gamma(a_\mathrm{s}\})$ are summarized in Table~\ref{table:beta-gamma}.
To facilitate the use of the Dy-DF for any value of the $a_s$ in the interval 
$a_s\in[60,110]$~$a_B$, the values 
$\{\beta(a_\mathrm{s}),\gamma(a_\mathrm{s}\})$ are interpolated in terms of 
$a_\mathrm{s}$ with a linear function. The parameters for these fits are 
summarized in Table~\ref{table:beta-gamma-fits} in the appendix. This interpolation allows to use the Dy-DF by introducing a minimal correction in LHY term of 
Eq.\eqref{eqn:MF-Efunc2}:  $\frac{2}{5}\Gamma_{\rm QF}|\Psi|^3 \rightarrow 
\beta_{\rm QMC}|\Psi|^{\gamma_{\rm QMC}+1}$, thus its implementation on any 
existing eGPE code is straightforward. It is worth remarking that the 
computational cost of the Dy-DF is exactly the same as the LHY-DF one. The 
quality of this approach is discussed in the next section.

\begin{figure}[tb]
    \centering
    \includegraphics[width=1.0\linewidth]{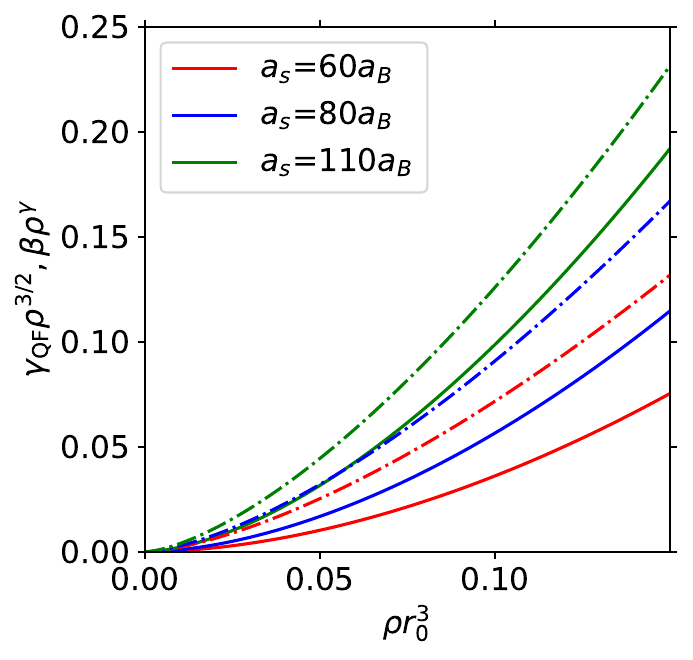} 
    \caption{Solid (dashed) lines are the quantum fluctuation term of the Dy-DF and the LHY-DF as a function of the density for different values of the scattering length. In the former case it corresponds to the non-linear term ($\beta\rho^\gamma$) of Eq.~\eqref{eqn:QMC-ekin-fit} and in the former to Eq.~\eqref{eqn:LHY}.}
    \label{fig:functionals}
\end{figure}

In Fig.~\ref{fig:functionals} we compare the quantum correlation term of the  Dy-DF with the one of the LHY-DF as a function of density. The values of the scattering length that are considered ($a_\mathrm{s} =$~60, 80 and 110~$a_\mathrm{B}$) cover the whole range of experimental values. As it can be observed in the figure, the Dy-DF is always less repulsive than the LHY one. Differences between the two functionals are larger when the scattering length is smaller what hints to a large beyond mean field correction when the dipolar interaction dominates.

\section{Critical atom number}
\label{sec:benchmarck}

\begin{figure}[t!]
    \centering
    \includegraphics[width=1.0\linewidth]{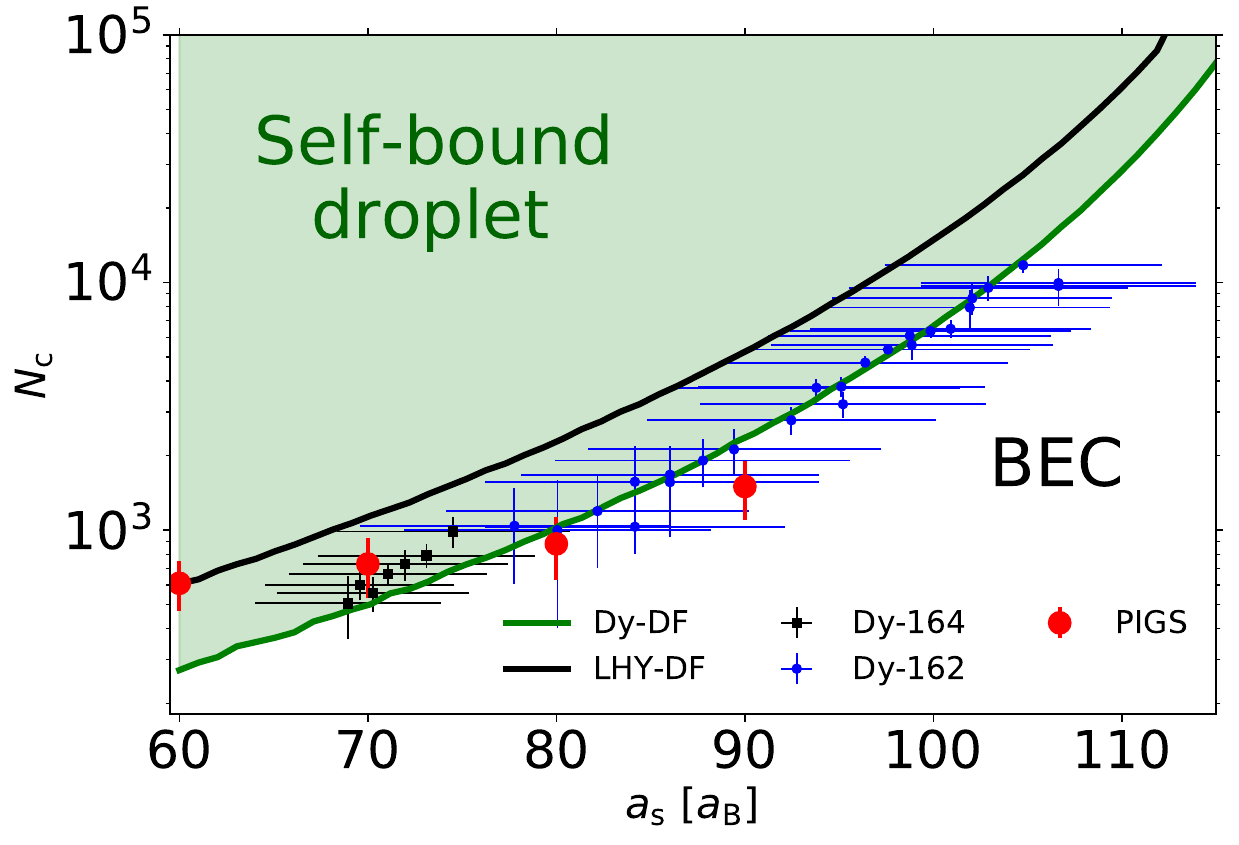}
    \caption{Critical atom number to form a dysprosium dipolar droplet. Green and black lines are Dy-DF and LHY-DF predictions, respectively.
    Black squares and blue dots are experimental measurements for $^{164}$Dy and $^{162}$Dy, respectively~\cite{Bottcher2019crit}. 
    Red points are obtained by direct calculations of quantum droplets with PIGS~\cite{Bottcher2019crit}. }
    \label{fig:critatomnum}
\end{figure}

\begin{figure*}
    \centering
    \includegraphics[width=1.0\linewidth]{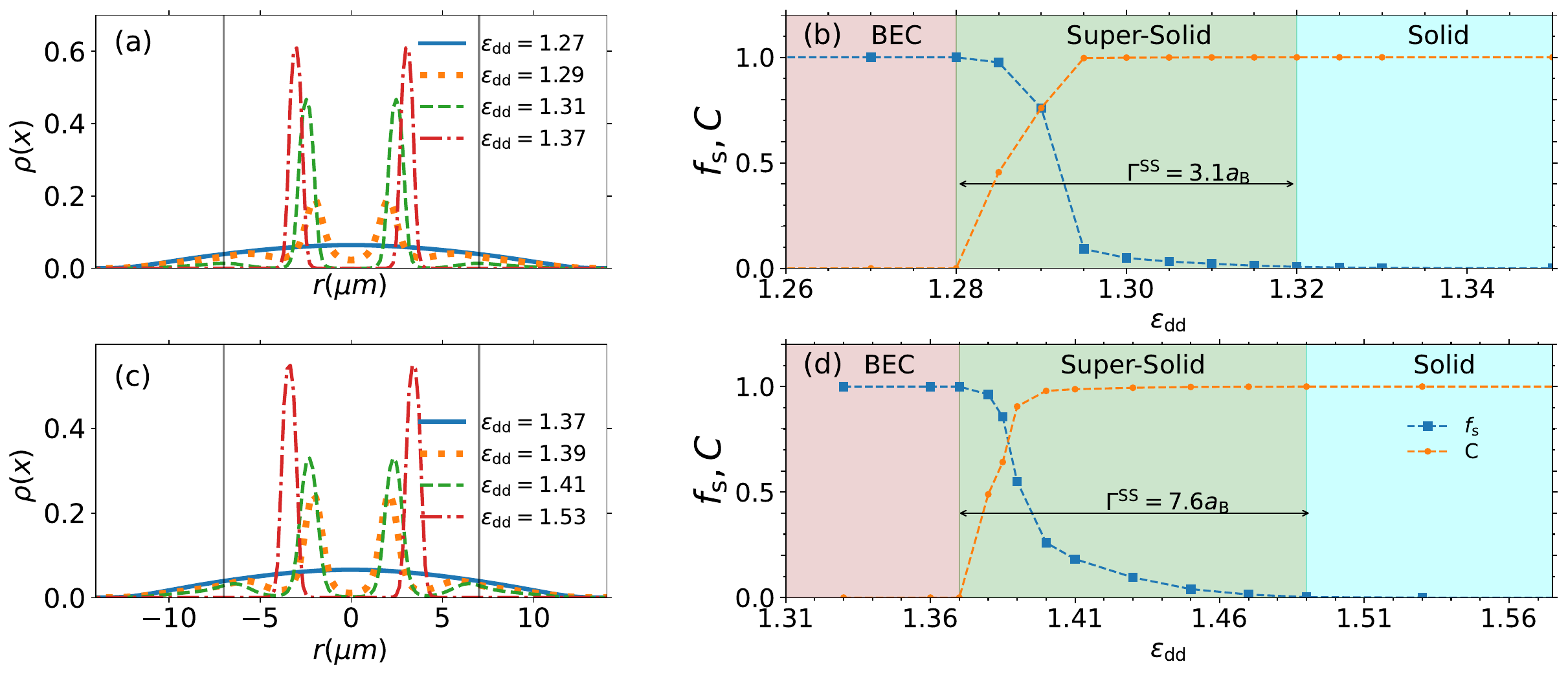} 
    \caption{Properties evaluated along the BEC-supersolid transition with the Dy-DF (top) and the LHY (bottom) functionals . (a) and (b) contain the density profiles along the $X$ direction for different values of $\varepsilon_\mathrm{dd}$. Panels (c) and (d) depict the superfluid fraction, $f_\mathrm{s}$, (blue squares, see Eq.~\eqref{eqn:legget}) and contrast, $C$,(orange circles, see Eq.~\eqref{eqn:contrast}) computed with the LHY-DF (left) and Dy-DF (right) as a function of $\varepsilon_\mathrm{dd}$.  Dashed lines are guides to the eye. Note that the scales in (b) and (d) are different. Grey vertical lines in (a) an (c) delimitate the region where $f_\mathrm{s}$ and $C$ are computed}
    \label{fig:transition}
\end{figure*}

The resulting Dy-DF is less repulsive than the LHY one, thus it is expected to predict a smaller critical atom number for droplet formation. As shown in Fig.~\ref{fig:critatomnum}, the Dy-DF functional (green line) reproduces with remarkable accuracy the available experimental data for the critical atom number of Dy atoms (black and blue dots). This improvement over the LHY-DF description (black line), which systematically overestimates this quantity, highlights the accuracy of the Dy-DF functional. Although our functional is designed for $^{162}$Dy, it also reproduces the experimental data for $^{164}$Dy, which suggests a minor isotopic effect in dysprosium droplets (note that the difference in mass and magnetic moment between the two atoms is about 1\%).

For the sake of comparison, we also include  in Fig.~\ref{fig:critatomnum} the PIGS results obtained by direct computation of small dipolar droplets~\cite{Bottcher2019crit}. PIGS results are compatible with the experimental values in the range $a_\mathrm{s} \in [70, 90]~a_\mathrm{B}$, where both PIGS calculations and experimental data are available. However, for $a_s = 60~a_\mathrm{B}$, the Dy-DF and the PIGS predictions are in clear disagreement. This difference can be attributed, at least in part, to the different methods used in both theories to estimate $N_\mathrm{c}$. With Dy-DF, $N_\mathrm{c}$ is calculated as the atom number for which the radius of the dipolar droplet starts to diverge in the absence of a trapping potential. This facilitates the direct comparison with experiments, where the same criteria to determine $N_\mathrm{c}$ is used. In the case of the direct PIGS calculations, using the size criteria is unfeasible. In that case, $N_\mathrm{c}$ was determined as the minimum atom number for which $E < 0$~\cite{Bottcher2019crit}. In appendix~\ref{app:crit-atom}, it is shown that the energy criterion provides critical atom numbers that are slightly larger than the ones obtained with the droplet radius criterion.

\section{BEC-supersolid transition}
\label{sec:BEC-supers-trans}

The BEC-supersolid transition, which arises in dipolar systems trapped in a quasi-one-dimensional configuration, has been extensively studied~\cite{Tanzi2019b,Bottcher2019,Natale2019,Chomaz2019,Guo2019,sohmen2021,Ilzhofer2021,Hertcorn2021,Tanzi2021}. Here, we focus on the experimental setup of Refs.~\cite{Tanzi2019a,Tanzi2019b,Bottcher2019}, where the trapped frequencies are $\omega_{x,y,z} = 2\pi(18.5,53,81)$~Hz, and the dipoles are aligned along the $Z$-axis with $N = 3.5\times10^{4}$ atoms. Similarly to what happens with droplet formation, the modulational instability that gives rise to the supersolid phase emerges from the balance between dipolar attraction and quantum correlations. Thus, the supersolid window, namely the interval of scattering length values at which this phase is observed, is quite sensitive to the details of the theoretical description. In this respect, as previously mentioned, the Dy-DF is more attractive than the usual LHY-DF, so the supersolid phase is expected to appear for larger values of the scattering length (lower $\varepsilon_{\rm dd}=a_{\rm dd}/a_s$).

Fig.~\ref{fig:transition}, (a) and (c), show the density profiles $\rho(x)$ 
along the $X$-direction, where the confinement is weaker,  calculated with  
Dy-DF and LHY-DF, respectively. The Dy-DF (LHY-DF) predicts a BEC state for $\varepsilon_{\mathrm{dd}}<1.27$ ($1.37$), while for larger values of $\varepsilon_\mathrm{dd}$, a density stability emerges, resulting in a modulated $\rho(x)$ distribution. Both functionals predict the formation of two large clusters in the center of the trap accompanied by two smaller satellite ones. As $\varepsilon_{\rm dd}$ is further increased, the intercluster density vanishes and a transition to a third regime occurs, where only two insulating droplets are present. This intermediate state is termed supersolid, as it exhibits spatial order and phase coherence simultaneously. Similarly, the system of insulating droplets constitutes a solid of droplets.

To characterize the transition between the three aforementioned regimes, namely BEC, supersolid, and solid, we evaluate the superfluid fraction, $f_s$, and the intensity of the density modulations. The Leggett's upper bound~\cite{Legget1970,Ghosh2022} for the superfluid fraction is given by
\begin{equation}
f_\mathbf{s} \le (2L)^2\left[\left(\int_{-L}^{L}dx\rho(x)\right)\left(\int_{-L}^{L}\frac{dx}{\rho(x)}\right)\right]^{-1}
    \label{eqn:legget}
\end{equation}
where the distance $2L$ encloses the central part of the trap, where the droplets form. Recently, the quality of Leggett's upper-bound has been studied, showing that it provides accurate results for dilute gases~\cite{Chauveau2023}. To evaluate the spatial structure of the modulated phases, we compute the contrast, $C$, between the height of the central peaks and the depletion of density around them as
\begin{equation}
    C = \frac{\mathrm{max}[\rho(x)]-\mathrm{min}[\rho(x)]}{\mathrm{max}
    [\rho(x)]+\mathrm{min}[\rho(x)]}.
    \label{eqn:contrast}
\end{equation}
The superfluid fraction and contrast are computed on the ground state predicted by the LHY and Dy-DF functionals. As we are dealing with a trapped system, we compute Eqs.~\eqref{eqn:legget} and ~\eqref{eqn:contrast} in the central region of the trap, which is delimited by the gray vertical lines in figures ~\ref{fig:transition}(a) and (c).

Figures \ref{fig:transition}(b) and (d) summarize the Dy-DF and LHY-DF results for the superfluid fraction and the contrast as a function of $\varepsilon_\mathrm{dd}$. The critical value of $\varepsilon_\mathrm{dd}$ at which the transition from a BEC to a supersolid phase occurs is $\varepsilon_\mathrm{dd}^\mathrm{crit}=1.28$ (1.38) for Dy-DF (LHY-DF). Below this critical value, $f_s = 1$ (and $C = 0$). In the supersolid region, the superfluid density rapidly drops to values lower than one, and the contrast increases. Note that values of $C$ close to one are obtained before the superfluid signal completely vanishes. For $\varepsilon_{dd}>1.32$ (1.48), Dy-DF (LHY-DF) predicts an insulating droplet array ($f_\mathrm{s}=0$ and $C = 1$).

\begin{figure}
    \centering    
    \includegraphics[width=1.0\linewidth]{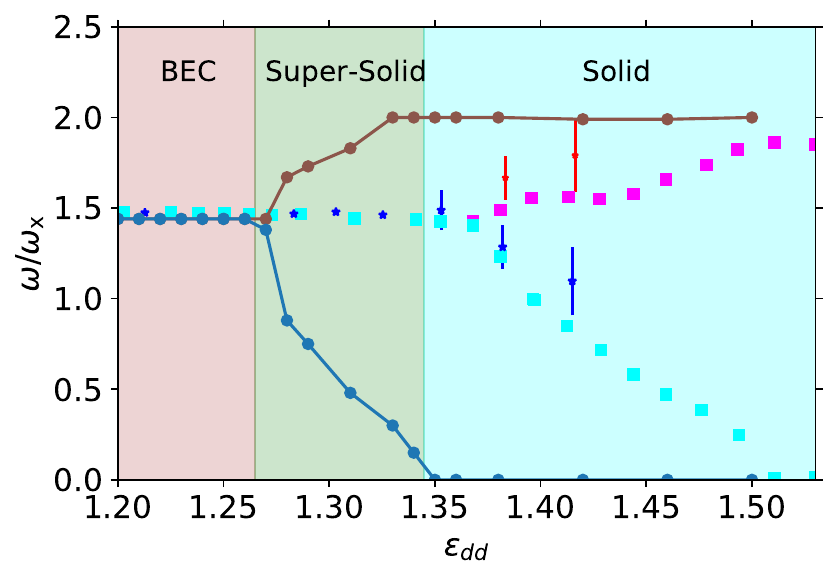}
    \caption{Axial mode frequencies in units of the trapping frequency $\omega_\mathrm{x}$ across the BEC to supersolid transition for the trap system of Fig.~\ref{fig:transition}. Circles connected by brown and blue lines are the Dy-DF predictions for the superfluid and solid frequencies. Blue and red points with error bars are the corresponding experimental measurements of Ref.~\cite{Tanzi2019b}. Pink and blue squares are LHY results from Ref.~\cite{Tanzi2019b}.}
    \label{fig:freqs}
\end{figure}

In Ref.~\cite{Tanzi2019b}, the axial mode frequencies were measured across the BEC-supersolid transition in the configuration that we study here. The authors showed that two modes can be excited in the supersolid regime. The high energy one is related to the lattice site vibrations, while the lower one is related to the droplet coherence oscillations. To compute the axial mode frequencies, we introduce a perturbative potential $H_{\rm pert} = -\lambda x^2$, with $\lambda$ being a small parameter. Once equilibrated, the perturbation is switched off, and the solution is propagated in real-time, computing $\left< x^2(t)\right>$. The axial mode frequencies are obtained by performing a Fourier transform of $\left< x^2(t)\right>$.

The axial mode frequencies computed with the Dy-DF are shown in Fig.~\ref{fig:freqs} (blue and brown circles). For $\varepsilon_{\rm dd}<1.27$, the system is in the BEC regime, and a single frequency is observed ($\omega/\omega_x \approx 1.47$). In the region $1.27<\varepsilon_{\rm dd}<1.34$, where the system is modulated (see Fig.~\ref{fig:transition}), two frequencies appear in the Fourier spectra. The magnitude of the higher energy mode is larger than the value of the single mode in the BEC regime, while the superfluid mode decreases quickly with $\varepsilon_{\rm dd}$ and vanishes for $\varepsilon_{\rm dd}>1.34$. This ending point signals the transition from the supersolid state to the normal solid one. In this phase, only the upper frequency branch appears with a value $\omega/\omega_x\approx 2.0$.

In Fig.~\ref{fig:freqs}, we also show the experimental measurement and LHY-DF prediction for these two frequencies as reported in Ref.~\cite{Tanzi2019b}. Note that the uncertainty in the measurement of $a_s$, which can be of the order of 10~$a_\mathrm{B}$, results in an uncertainty of 0.12 in the value of $\varepsilon_{dd}$. More refined experimental measurements would be needed in order to 
determine precisely the supersolid transition point and to discern between the 
Dy-DF and LHY-DF predictions.
Despite of that, let us analyze the differences between the two functionals.
In the BEC regime both functionals reproduce the experimental value 
$\omega_x/\omega \approx1.47$ within a $3\%$ of error. In the supersolid regime 
the behavior the LHY-DF modes is similar  to that of the 
Dy-DF but it is noticeable that the interval in which the two frequencies 
appear is larger in the former case. 
The LHY superfluid frequency is close to the experimental one but it underestimates the value of the lattice frequency. 
On the contrary, the Dy-DF prediction reproduces the frequency of the
experimental lattice mode but underestimates the superfluid one.
In the solid regimes both functionals predict a single frequency mode at  $\omega/\omega_x= 2.00$,~1.85 in for Dy-DF and LHY-DF, respectively.

An important difference between the two theoretical approaches is the difference 
in the width of the supersolid window in terms of the scattering length, 
$\Gamma^{\mathrm{SS}}$. The experimental estimation of the supersolid window for 
this same setup  given in Ref.~\cite{Bottcher2019} was obtained by measuring in-situ 
modulations and phase coherence between the different drops. 
The reported value is $\Gamma^{\mathrm{SS}}_{\rm exp}\approx 6~a_\mathrm{B}$. 
It is worth noticing that both for Dy-DF and LHY-DF functionals the prediction 
for $\Gamma^{\mathrm{SS}}$ is rather different if it is evaluated 
in terms of the values of $f_s$ of Fig.~\ref{fig:transition} or of the 
excitation spectra of Fig.~\ref{fig:freqs}. The second choice provides slightly 
larger values.
The LHY-DF prediction for the supersolid window attending to $f_\mathrm{s}$ 
is $\Gamma^{\mathrm{SS}}_{\rm LHY}\approx 7.6~a_\mathrm{B}$ while the 
width of the interval at which two frequencies appear in the Fourier spectra of 
Fig.~\ref{fig:freqs} is of 8.8~$a_\mathrm{B}$. In the case of the Dy-DF 
functional, the first (second) criteria gives $\Gamma^{\mathrm{SS}}_{\rm 
Dy-DF}\approx 3.1 (5.3)~a_B$. Thus, the experimental measurement of 
Ref.~\cite{Bottcher2019} lies in between the results of the two 
functionals.

\section{Discussion and conclusions.}
\label{sec:conclusions}

Summarizing, the Dysprosium density functional (Dy-DF) that we present in this work allows to study Dy dipolar systems and the rich phenomena that they exhibit, such as droplet formation and supersolidity. We evaluate the properties of a bulk dipolar system made of $^{162}$Dy atoms using first-principles quantum Monte Carlo. The functional is constructed under the local density approximation by replacing the usual Lee-Huang-Yang (LHY) term, which accounts for quantum correlations in the standard extended Gross-Pitaevskii equation (eGPE), with exact quantum correlations computed with QMC. To benchmark our functional, we compute the minimal (critical) atom number that is needed to form a droplet and show that our theory reproduces the experimental measurements for this quantity in spite of the large uncertainties in the experimental determination of the scattering lengths. The critical atom number is a challenging observable from a theoretical point of view, as it arises from a delicate energy balance between two similar (and opposite sign) quantities: the interatomic dipolar interaction and the contribution from quantum correlations. On the contrary the LHY theory only provides a qualitative description of the droplet formation mechanism. Furthermore, we demonstrate the suitability of our functional to study the BEC to supersolid phase transition and discuss the small discrepancies that appear between the LHY theory and our functional. Importantly, the Dy-DF functional improves the accuracy of the widely used eGPE without increasing the computational cost.  
A lot of work have been done in the last years to achieve condensates of dipolar molecules that have a large electric moment~\cite{Lin2023,Bigagli2023,Schmidt2022,Mukherjee2023} and for which the intermolecular interaction is known with spectroscopic precision~\cite{Deng2023}. Recently, the formation of a Bose-Einstein condensate of dipolar molecules have been reported~\cite{bigagli2023cond}, what constitues a breaktrhough that opens the door to study new physics in dipolar systems. We hope that the approach presented here will be useful also for the study of such condensates of dipolar molecules in the future

\acknowledgments{
We acknowledge Tilman Pfau and Fabian B\"ottcher for sharing with us the experimental data set for the critical atom number.
We acknowledge financial support from Ministerio de Ciencia e Innovaci\'on 
MCIN/AEI/10.13039/501100011033
(Spain) under Grant No. PID2020-113565GB-C21 and
from AGAUR-Generalitat de Catalunya Grant No. 2021-SGR-01411. 
R.B. acknowledges European  Union-NextGenerationEU, Ministry of Universities and Recovery, Transformation and Resilience Plan, through a call from Polytechnic University of Catalonia.
R.B. further acknowledges funding from ADAGIO (Advanced ManufacturIng Research Fellowship Programme in the Basque – New Aquitaine Region)  MSCA COFUND post-doctoral fellowship programme.
}

\appendix

\section{Fitting procedure}
\label{app:PIGSfit}

\begin{table}[t!]
    \caption{Parameters for the quantum fluctuation term in the energy per particle at LHY-DF level and as extracted from  Eq.~\eqref{eqn:QMC-ekin-fit} to construct the Dy-DF. In the LHY case $\beta_{\rm LHY} = \frac{2}{5}\Gamma_{\rm QF}$ and $\gamma_{LHY} = 1.5$ (see Eq.~\eqref{eqn:mu}). The Dy-DF values are obtained after averaging different fits where the fitting interval of the PIGS data is slightly varied and the number in parenthesis corresponds to standard deviation, see text for details.}
    \label{table:beta-gamma}
\begin{ruledtabular}
\begin{tabular}{ c c c c c  }
 & \multicolumn{2}{c}{LHY-DF} & \multicolumn{2}{c}{Dy-DF}\\\cline{2-3} \cline{4-5}
$a_s/a_B$& $\beta=\frac{2}{5}\Gamma_{\rm QF} [\mathcal{E}_0r_0^3]$ & $\gamma$ & $\beta [\mathcal{E}_0r_0^3]$ & $\gamma$\\
\hline
60  & 2.269 & 1.5 & 3.0(3) & 2.00(6) \\
70  & 2.562 & 1.5 & 1.93(15) & 1.53(7) \\
80  & 2.876  & 1.5 &3.35(8) &  1.80(1)\\
90  & 3.215 & 1.5 & 3.3(9) &  1.65(13)\\
101.57 & 3.583  & 1.5 &3.4(5) &  1.62(7)\\
110 & 3.984  & 1.5 & 5.0(7) & 1.73(6) \\
\end{tabular}
\end{ruledtabular}
\end{table}

\begin{table}[tb]
    \centering
    \caption{The parameter $\beta$ and $\gamma$ of Table~\ref{table:beta-gamma} are interpolated in terms of the scattering length, $a_\mathrm{s}$ with a linear function $ma_s + n$. The values of $m$ and $n$ for the two parameters are given in the table.}
    \label{table:beta-gamma-fits}
\begin{ruledtabular}
        \begin{tabular}{ccc}
        $\beta$ & $m [\mathcal{E}_0r_0^2]$ & $n [\mathcal{E}_0r_0^2]$ \\\cline{1-1}\cline{2-3}
         & 0.03935 & -0.01422 \\\\
     $\gamma$    & $m [r_0^{-1}]$ & $n$ \\\cline{1-1}\cline{2-3}
        & -0.003434  &2.0182 \\
    \end{tabular}
\end{ruledtabular}
\end{table}

The Dy-DF is constructed by replacing the LHY term, accounting for quantum correlations in the eGPE description, by the non-linear contribution of the quantum kinetic energy, namely, quantum correlations.
Although QMC provides us with exact results for bosonic systems, they are accompanied by a certain variance due to the stochastic nature of the method. 
To deal with it we make linear fits of $\left<\hat{T}_{\phi}\right>_0$ to a function of the form $\alpha\rho + \beta\rho^\gamma$, in different intervals of the density, $[\rho_{\textrm{min}}, \rho_\textrm{max}]$, around the equilibrium density $\rho_\mathrm{eq}$ (see Table.~\ref{table:eq-dens}). 
We consider all the possible combinations of $\rho_{\textrm{min}}$ and $ \rho_\textrm{max} $ such as $\rho_{\mathrm{min}}/\rho_{\mathrm{eq}}\in[0.4,0.7]$ and $\rho_{\mathrm{max}}/\rho_{\mathrm{eq}}\in[1.5,2.5]$, considering increments of 0.1 and 0.3, respectively.
The resulting mean values of $\beta$ and $\gamma$  are listed din Table~\ref{table:beta-gamma} with their variance indicated in parenthesis. Code to reproduce the fitting procedure is given here \cite{cikojevic_raul2023}. Note that the parameter $\alpha$ in Eq.~\eqref{eqn:QMC-ekin-fit} is not used for the construction of the functional and thus is not given on the table to avoid confusions. The values of $\alpha$ in the fit of the kinetic energy are 
249(6), 760(13), 924(15), 2818(695), 8715(1537), 21185(3407) for $a_\mathrm{s}= 60, 70, 80, 90, 101, 110$~a$_\mathrm{B}$, respectively.

We perform a linear interpolation of the  values of  $\beta$ and $\gamma$ in Table~\ref{table:beta-gamma} to create a functional dependence on $ a_s $. The parameters for the linear fits of $\beta$ and $\gamma$ in terms of $a_s$ are given in table~\ref{table:beta-gamma-fits}. Code to reproduce the fitting procedure is given here \cite{cikojevic_raul2023}. This procedure allows to use the Dy-DF  for any value of the scattering length in the interval $[60,110]~a_\mathrm{B}$. It also allows to implement straightforwardly the Dy-DF functional in existing eGPE code.

\section{Evaluation of critical atom number}
\label{app:crit-atom}

\begin{figure*}
    \centering
    \includegraphics[width=\linewidth]{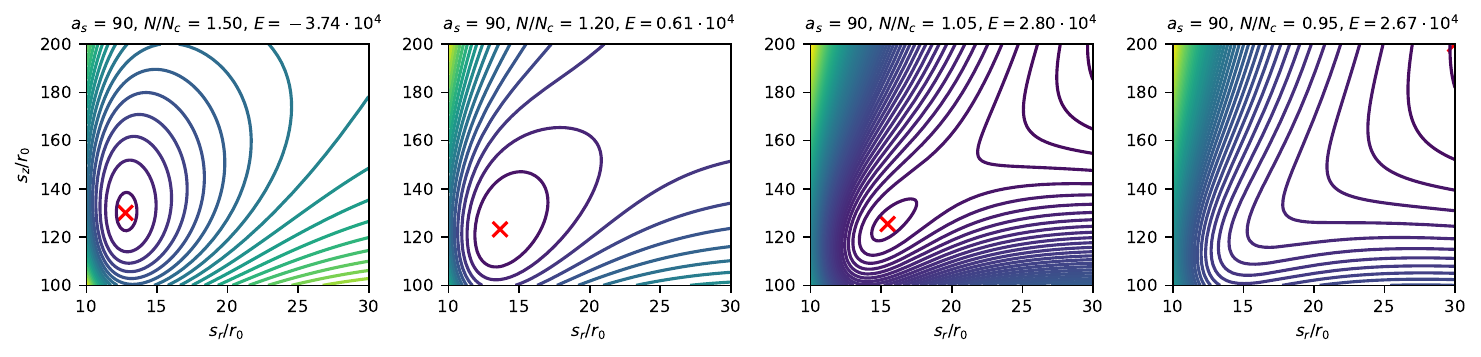}
    \caption{Energy landscape in terms of the Gaussian widths $\sigma_z$ and $\sigma_r$, for $a_s = 90 a_0$ and various \( N/N_c \) ratios. The red marker indicates local minima. The values of the $N/N_\mathrm{c}$ ratios an the energy of the local minima are indicated on each panel.}
    \label{fig:energy-landscape-appendix-b}
\end{figure*}

We evaluate the critical atom number within the density functional formalism. To this end, we employ a Gaussian ansatz parametrized in the radial and axial directions with the widths \( s_r \) and \( s_z \), respectively. The normalized density of a given Gaussian wave function is given by:

\begin{equation}
    \label{appendix-b-ansatz}
    \rho(r,z) = \frac{N}{\sqrt{\pi s_r^2 s_z}} e^{-\frac{r^2}{2s_r^2} - \frac{z^2}{2s_z^2}} 
\end{equation}
where \( N \) is the total number of atoms. Withing this ansatz, the energy of the system can be calculated after plugging it into Eq. \ref{eqn:MF-Efunc2}. We define the critical atom number as the number of atoms under which the local minimum of the energy landscape \( E(s_r, s_z) \) vanish. Notice that the energy can be positive, signaling a metastable state. 
To give further insight into this, in Fig. \ref{fig:energy-landscape-appendix-b}, we plot the energy landscape for \( a_s = 90~a_0 \) for various \( N/N_c \) ratios. For this purpose, the Dy-DF parameters taken from table \ref{table:beta-gamma-fits} are used. It is evident that the energy of a local minimum decreases with a decrease in the \( N/N_c \). Notably, at around \( N/N_c = 1.2 \), the energy becomes positive, meaning that the global minimum vanishes for finite \( s_r \) and \( s_z \) values. Despite this, a local minimum persists, indicating the presence of a metastable minimum. This metastable state ceases to exist when \( N/N_c \) drops below one. The red markers on the figure serve to highlight the points of local minima. Obviously, for $N/N_\mathrm{c}<1$ no local minima is observed.

\bibliography{apssamp}

\end{document}